\documentclass[aps,pra,amsmath,amssymb, reprint,superscriptaddress, floatfix]{revtex4-2}
\newcommand{\bol}[1]{\boldsymbol{#1}}
\usepackage{physics}
\usepackage{calc}
\usepackage{mathtools}
\newcommand{\cmmnt}[1]{}
\usepackage{tikz}
\usetikzlibrary{quantikz2}
\usepackage{enumitem}
\usepackage{balance}
\usepackage[pdftex,pdfpagelabels,bookmarks,hyperindex,hyperfigures,colorlinks]{hyperref}
\begin{document}

%\preprint{}

\title{Quantum implementation of non-unitary operations with biorthogonal representations}


\author{Efstratios Koukoutsis}
\email{stkoukoutsis@mail.ntua.gr}
\author{Panagiotis Papagiannis}
\author{Kyriakos Hizanidis}
\affiliation{School of Electrical and Computer Engineering, National Technical University of Athens, Zographou 15780, Greece}
%\affiliation{School of Electrical and Computer Engineering, National Technical University of Athens, Zographou 15780, Greece}
\author{Abhay K. Ram}
\affiliation{Plasma Science and Fusion Center, Massachusetts Institute of Technology, Cambridge,
Massachusetts 02139, USA}
\author{George Vahala}
\affiliation{Department of Physics, William \& Mary, Williamsburg, Virginia 23187, USA}
\author{{\'O}scar Amaro}
\author{Lucas I I{\~n}igo Gamiz}
\affiliation{GoLP/Instituto de Plasmas e Fus{\~a}o Nuclear, Instituto Superior T{\'e}cnico, Universidade de Lisboa,
1049-001 Lisbon, Portugal}
\author{Dimosthenis Vallis}
\affiliation{School of Electrical and Computer Engineering, National Technical University of Athens, Zographou 15780, Greece}
%homepage[jfbjdjdb]{Your web page}
%\thanks{}
%\altaffiliation{λκξλξλξλξ}


\date{\today}

\begin{abstract}
Motivated by the contemporary advances in quantum implementation of non-unitary operations, we propose a new dilation method based on the biorthogonal representation of the non-unitary operator, mapping it to an isomorphic unitary matrix in the orthonormal computational basis. The proposed method excels in implementing non-unitary operators whose eigenvalues have absolute values exceeding one, when compared to other dilation and decomposition techniques. Unlike the Linear Combination of Unitaries (LCU) method, which becomes less efficient as the number of unitary summands grows, the proposed technique is optimal for small-dimensional non-unitary operators regardless of the number of unitary summands. Thus, it can complement the LCU method for implementing general non-unitary operators arising in positive only open quantum systems and pseudo-Hermitian systems.
\end{abstract}

%\keywords{}

\maketitle

\section{Introduction}\label{sec:1}
Implementing non-unitary operations in a quantum computer is an inherently challenging task due to the unitary operational framework of the quantum machines. However, non-unitarity resides in a realm of exceptional physical interest accommodating a plethora of quantum technology advances in areas such as open quantum systems \cite{Breuer_2007,Delgado_2024}, pseudo-- and PT-- Hermitian systems \cite{Mostafazadeh_2010,Ju_2019} and quantum simulation of differential equations\cite{Childs_2012,Berry_2015,Krovi_2023}.  

\label{eq1}
Contemporary methods on treating non-unitary quantum gates rely on duality quantum computing \cite{Long_2006,Zheng_2021} and subsequently to the Linear Combination of Unitary (LCU) method \cite{Childs_2012} and other unitary decompositions \cite{Schlimgen_2021,Suri_2023} as well as to dilation methods \cite{Schlimgen_2022,Schlimgen2_2022,Jin_2023,Koukoutsis_2024}. In principle, both techniques require the introduction of ancillary qubits to mimic an appropriate environment where the non-unitary operators obtain a unitary representation in an enlarged Hilbert space \cite{Moshe_2021}. 

In the spirit of the aforementioned research, this paper aims to establish a novel dilation protocol, inspired by the biorthogonal quantum mechanics \cite{Brody_2013}, for a specific class of non-unitary operators that possess a unitary counterpart in the biorthogonal representation.

In  standard quantum computing, states and operations are implemented under a  complete and orthonormal  basis set $\{\ket{\bol k}\}$ provided by a Hermitian operator $\hat{O}$--observable. However,  when $\hat O$ is non-Hermitian  there is a set $\{\ket{\bol u}, \ket{\bol\zeta}\}$ composed by the right and left eigenvectors of  $\hat O$,
\begin{align}
\hat{O}\ket*{u_n}=\lambda_n\ket*{u_n},& \quad \bra*{u_n}\hat{O}^{\dagger}=\lambda^*_n\bra*{u_n}, \label{eq2} \\ 
\hat{O}^{\dagger}\ket*{\zeta_n}=\lambda^*_n\ket*{\zeta_n},&\quad \bra*{\zeta_n}\hat{O}=\lambda_n\bra*{\zeta_n},\label{eq3}
\end{align}
that forms a biorthogonal basis with an overlapping orthogonality relation,
\begin{equation}\label{eq4}
\braket*{\zeta_n}{ u_m}=\braket*{\zeta_n}{ u_n}\delta_{nm},\quad\text{(no summation implied)}.
\end{equation}

By treating the biorthogonal basis  set $\{\ket{\bol u}, \ket{\bol\zeta}\}$ as a new quantum representation for the $\mathcal{N}$--qubit non-unitary operator $\hat{V}$ we devise a proper biorthogonal counterpart $\hat{V}^b$ that attains a unitary matrix representation in the computational basis $\{\ket{\bol k}\}$. This unitary representation $\hat{\mathcal{V}}$, mediates the post-selective implementation of the non unitary action $\hat{V}\ket{\Psi}$ in a dilated space consisting of $\mathcal{N}$ ancillary qubits.

In contrast to the LCU method, the quantum implementation scaling of the biorthogonal dilation is insensitive to the number of unitary operators in which the non-unitary operator is decomposed. This attribute, makes the biorthogonal dilation technique particularly effective in quantum implementation of non-contraction non-unitary operators of small dimension. Thus, an effective scheme incorporating both the LCU and the biorthogonal method for implementing complex non-unitary operations is proposed with applications to positive only quantum channels and pseudo-Hermitian systems.

The structure of the paper is as follows: Section \ref{sec:2} serves as an exposition of the basic elements and definitions of the biorthogonal framework, introducing the associated states and the biorthogonal operations in an analogy with the standard quantum mechanics. In Sec.\ref{sec:3.1} the basic implementation algorithm for the non-unitary operator $\hat{V}$ is presented under the condition that its biorthogonal representation $\hat{V}^b$ is biorthogonaly unitary (bi-unitary). Then, in Sec.\ref{sec:3.2} the respective implementation steps of the algorithm are demonstrated for a one-qubit non-unitary matrix. Section \ref{sec:3.3} relaxes the bi-unitarity condition, extending the applicability domain of the proposed dilation method to arbitrary non-unitary matrices beyond contractions.  In Sec.\ref{sec:3.4} a direct comparison with the Linear Combination of Unitaries (LCU) method and Sz.-Nagy dilation is performed, in terms of efficiency for the different classes of non-unitary operators. Finally, Sec.\ref{sec:3.5} considers a range of applications where the associated non-unitary operators possess complex stability features, for which the biorthogonal dilation in conjunction with the LCU method could facilitate an efficient quantum implementation.

\section{The biorthogonal framework}\label{sec:2}
Following the seminal work in \cite{Brody_2013}, we present the building blocks of the biorthogonal quantum mechanics, adding the necessary elements pertinent to our purposes. Throughout the paper the involved Hilbert spaces are considered to be finite dimensional and the indices $n,m,k$ span $\{0,1,...,2^{\mathcal{N}}-1\}$.

\subsection{Associated spaces, states and inner product}\label{sec:2.1}
 Any state $\ket*{\psi}\in\mathcal{H}$ can be written in terms of a non-orthogonal basis set  $\{\ket*{\bol u}\}$ that spans the Hilbert space $\mathcal{H}$ as,
\begin{equation}\label{eq5}
\ket*{\psi}=\sum_{n}c_n\ket*{u_n},\quad \braket{ u_m}{ u_n}\neq\delta_{nm},
\end{equation}
where the respective bra set $\{\bra*{\bol u}\}$ belongs to the dual space $\mathcal{H}^*$. To obtain a biorthogonal set, we define a linear, invertible mapping $\bol{\hat{f}}$ that projects the non-orthogonal basis elements $\{\ket{\bol{u}}\}$ into their respective orthogonal elements in the Hilbert space $\tilde{\mathcal{H}}$, forming a new non-orthogonal basis set $\{\ket{\bol\zeta}\}$,
\begin{equation}\label{eq6}
\bol{\hat{f}}:\mathcal{H}\to\tilde{\mathcal{H}},\quad \bol{\hat{f}}\ket*{\bol u}=\ket*{\bol \zeta},\quad \{\ket{\bol{u}} \perp \ket{\bol\zeta}\}.
\end{equation}
The Hilbert space $\tilde{\mathcal{H}}$ is the associated state space. The action of $\bol{\hat f}$ and $\bol{\hat f}^{-1}$ mappings as orthogonal projections between the Hilbert spaces, $\mathcal{H}$ and its associated $\tilde{\mathcal{H}}$, is explicitly provided by
\begin{equation}\label{eq20}
\bol{\hat f}=\sum_n\frac{1}{\kappa_n}\ket{\zeta_n}\bra{\zeta_n},\quad \bol{\hat f}^{-1}=\sum_n\frac{1}{\kappa_n}\ket{u_n}\bra{u_n},
\end{equation}
with  $\kappa_n=\braket*{\zeta_n}{u_n}>0$. By definition, the mapping $\bol{\hat f}$ in Eq.\eqref{eq20} is Hermitian and naturally introduces the associate state of $\ket*{\psi}$,
\begin{equation}\label{eq7}
\ket*{\tilde{\psi}}=\bol{\hat{f}}\ket*\psi=\sum_n c_n \hat{f}_n\ket*{u_n}=\sum_n c_n\ket*{\zeta_n}\in\tilde{\mathcal{H}}.
\end{equation}
Accordingly, the bra-associated state $\bra*{\tilde{\psi}}$ given by
\begin{equation}\label{eq8}
\bra*{\tilde{\psi}}=(\ket*{\tilde{\psi}})^\dagger=\sum_n\bra*{\zeta_n}c^*_n,
\end{equation}
belongs to the dual associated space $\bra*{\tilde{\psi}}\in\tilde{\mathcal{H}}^*$.

Definitions \eqref{eq5} and \eqref{eq7} suggest that the bi-linear functional $\braket{}{}:\tilde{\mathcal{H}}\times\mathcal{H}\to\mathbb{C}$, constitutes a proper inner product,
\begin{equation}\label{eq9}
\braket*{\tilde{\phi}}{\psi}=\sum_{n,m}d_n^* c_m\braket*{\zeta_n}{u_m}=\sum_n \kappa_nd_n^* c_n.
\end{equation}
Thus, any  state $\ket*{\psi}$ can be normalized in terms of the biorthogonal norm \eqref{eq9} as in standard quantum mechanics,
\begin{equation}\label{eq10}
\ket*{\psi}\to\frac{\ket*{\psi}}{\sqrt{\braket*{\tilde{\psi}}{\psi}}}=\frac{\sum_n c_n\ket*{u_n}}{\sqrt{\sum_n \kappa_n\abs{c_n}^2}}.
\end{equation}

The inner product \eqref{eq9} also suggests the definition of a new involution operation $^\ddagger$ such that,
\begin{equation}\label{eq11}
(\ket*{\psi})^\ddagger=\bra*{\tilde{\psi}},
\end{equation}
that is called the biorthogonal complex conjugation. The utility  of operation \eqref{eq11} lies in offering a new involution for characterizing operator properties within the biorthogonal framework, complementing the standard complex conjugation $^\dagger$ involution. Thus, an operator $\hat{V}$ in the orthonormal basis $\ket{\bol k}$ may possess attractive properties under the $^\ddagger$ involution in the biorthogonal representation $\hat V^b$. This in turn makes the new involution particularly useful for translating operator properties between the two representations.

\subsection{Operators}\label{sec:2.2}
In the established biorthogonal framework any operator $\hat V^b$ has an outer product form,
\begin{equation}\label{eq12}
\hat{V}^b=\sum_{n,m} V_{nm}\ket*{u_n}\bra*{\zeta_m},
\end{equation}
where $V_{nm}$ is the  matrix representation of operator $\hat{V}^b$, and it is isomorphic to the operator $\hat{\mathcal{V}}$,
\begin{equation}\label{eqextra1}
V_{nm}\leftrightarrow\hat{\mathcal{V}}=\sum_{n,m}V_{nm}\ket{n}\bra{m},
\end{equation}
expressed in the orthonormal basis $\{\ket{\bol k}\}$.
Immediately, from Eq.\eqref{eq12} the completeness relation reads,
\begin{equation}\label{eq13}
\hat{1}^b=\sum_{n} \frac{1}{\kappa_n}\ket*{u_n}\bra*{\zeta_n},
\end{equation}
with $\hat{1}^b$ being the biorthogonal unity operator, $\hat{1}^b\ket*{\psi}=\ket*{\psi}$.

In the following, the translation of the Hermicity and unitarity notion in the standard quantum mechanics into the biorthogonal extension is presented.

\subsubsection{Biorthogonal Hermitian operators}
In quantum mechanics, the set of physically meaningful measurement outcomes, known as observables, corresponds to a set of Hermitian operators ensuring the reality of the outcome,
\begin{equation}\label{eq16}
\mel{\Psi}{\hat V}{\Psi}\in\mathbb{R}\Leftrightarrow \hat{V}=\hat{V}^{\dagger}.
\end{equation}
In the biorthogonal extension, the reality of observables is secured under an analogous condition with Eq.\eqref{eq16}, but now the associated operator has to be biorthogonaly Hermitian (bi-Hermitian), i.e. Hermitian under the new involution defined in Eq.\eqref{eq11},
\begin{equation}\label{eq17}
\mel*{\tilde{\psi}}{\hat{V}^b}{\psi}\in\mathbb{R}\Leftrightarrow \hat{V}^b=\hat{V}^{b^\ddagger},
\end{equation}
with the state $\ket*{\psi}$ properly normalized according to Eq.\eqref{eq10}.

By explicitly calculating the mean value quantity in Eq.\eqref{eq17}. using Eqs.\eqref{eq5},\eqref{eq8} and \eqref{eq12} we obtain that
\begin{equation}\label{eq18}
\hat{V}^b=\hat{V}^{b^\ddagger} \Leftrightarrow V_{nm}={V^*_{mn}}.
\end{equation}
In addition, the standard complex conjugation involution and its biorthogonal counterpart are related through
\begin{equation}\label{eq19}
\hat{V}^{b^\dagger}=\bol{\hat f}\hat{V}^{b^\ddagger}\bol{\hat f}^{-1}.
\end{equation}

Equation \eqref{eq19}  shows that a bi-Hermitian operator, $\hat{V}^b=\hat{V}^{b^\ddagger}$, possesses a pseudo-Hermitian structure \cite{Mostafazadeh_2002,Mostafazadeh_2010,Znozil_2008} with the Hermitian  operator $\bol{\hat{f}}$ acting as the metric operator. If a positive definite metric operator $\bol{\hat{f}}^+$, from the family of $\{\bol{\hat{f}}\}$  exists \cite{Mostafazadeh_2003}, a  Dyson map $\hat{\eta}$, $\bol{\hat{f}}^+=\hat{\eta}^\dagger\hat{\eta}$ can be constructed. The Dyson map establishes an equivalence between Hermicity and bi-Hermicity in the Hilbert space $\mathcal{H}$, facilitating quantum computing tasks using the $\hat{\eta}\ket*{\psi}$ states \cite{Koukoutsis_2023}.

\subsubsection{Biorthogonal unitary operators}
In analogy with the definition of unitary operators in the orthonormal case, a biorthogonal unitary (bi-unitary) operator, $\hat{V}^b$, preserves the inner product structure of Eq.\eqref{eq9},
\begin{equation}\label{eq21}
\mel*{\tilde{\phi}}{\hat{V}^{b^\ddagger}\hat{V}^b}{\psi}=\braket*{\tilde{\phi}}{\psi}\Leftrightarrow \hat{V}^{b^\ddagger}\hat{V}^b=\hat{1}^b.
\end{equation}
With the aid of Eq.\eqref{eq19}, the bi-unitarity condition \eqref{eq21} in terms of the the complex conjugation $^\dagger$ involution reads,
\begin{equation}\label{eq22}
\hat{V}^{b^\dagger}\bol{\hat f}\hat{V}^b=\bol{\hat f}.
\end{equation}
 
Equation \eqref{eq22} reveals that under the usual complex conjugation, bi-unitary operators possess a complex Lorentz transformation structure \cite{Zhang_2018} with metric $\bol{\hat f}$. Hence, a biorthogonal unitary operator that preserves the proper biorthogonal inner product of Eq.\eqref{eq21}  also constitutes a Lorentz transformation in the complex Hilbert space $\mathcal{H}$, preserving the generally indefinite inner product $\mel{\psi}{\bol{\hat f}}{\psi}$.

\section{Implementing non-unitary operators with biorthogonal unitary representations}\label{sec:3}
Throughout this section we fix $\norm{u_n}^2=1$ to resolve the rescaling ambiguity in the biorthogonal vectors \cite{Edvardsson_2023}. In addition, we temporarily set $\kappa_n=1$. Then, according to Eq.\eqref{eqextra1}, the action of the biorthogonal operators is equivalent to the action of its matrix representation in the orthonormal basis,
\begin{equation}\label{eq23}
\hat{V}^b\leftrightarrow \hat{\mathcal{V}},\quad \ket{\bol u}\leftrightarrow \ket{\bol k}.
\end{equation}
Therefore, the $\hat{V}^b\ket{\psi}$ action reads,
\begin{equation}\label{eq24}
\begin{split}
\hat{V}\ket{\Psi}&=\hat{V}^b\ket{\psi}=\sum_{n,m}V_{nm}c_m\ket{u_n}=\sum_{n,m}V_{nm}c_m\hat{U}_n\ket{n}\\
=&\sum_n \hat{U}_n(\hat{\mathcal{V}}\ket{\Bar{\psi}})_n.
\end{split}
\end{equation}
In Eq.\eqref{eq24}, the $(\,\,)_n$ notation means the $n$-th component of the state inside the parenthesis. Also, the various states and operators in Eq.\eqref{eq24} in terms of the orthonormal and biorthonormal bases read,
\begin{alignat}{2}
&\hat{V}=\sum_{n,m}v_{nm}\ket{n}\bra{m},\quad &&\ket{\Psi}=\sum_k a_k\ket{k},\label{eq25} \\ 
&\hat{\mathcal{V}}=\sum_{n,m} V_{nm}\ket{n}\bra{m},\quad &&\ket{\Bar{\psi}}=\sum_kc_k\ket{k}. \label{eq26}
\end{alignat}
Finally, the condition $\norm{u_n}^2=1$ dictates that the $\hat{U}_n$ operators are unitary operators, mapping each element from the orthonormal basis set to the respective non-orthogonal basis, $\hat{U}_n\ket{n}=\ket{u_n}$.

The interconnection relationship in Eq.$\eqref{eq24}$ indicates that when the operator $\hat{\mathcal{V}}$ is unitary, the non-unitary action $\hat{V}\hat{\Psi}$ can be realized as a dot-product operation of unitary components $(\hat{U}_n, \hat{\mathcal{V}})$ in the $\ket{\Bar{\psi}}$ state, expressed within the computational basis $\{\ket{\bol k}\}$.

From the definition of bi-unitarity Eq.\eqref{eq21} for $\kappa_n=1$, and Eq.\eqref{eqextra1} the matrix $\hat{\mathcal{V}}$ is unitary if and only if $\hat{V}^b$ is bi-unitary. Consequently, implementing the non-unitary operation $\hat{V}$ in a quantum computer can be realized through its biorthogonal  representation $\hat{V}^b$, provided that  $\hat{V}^b$ is bi-unitary. In turn, the bi-unitary structure of $\hat V^b$ entails the unitarity of operation $\hat{\mathcal{V}}$ which carries the implementation of $\hat{V}$ in the computational basis $\{\ket{\bol{k}}\}$ as per Eq.\eqref{eq24},
\begin{equation}\label{balader}
\braket{\Bar{\psi}}{\Bar{\psi}}=\mel{\Bar{\psi}}{\hat{\mathcal{V}}^\dagger\hat{\mathcal{V}}}{\Bar{\psi}}=\mel*{\tilde{\psi}}{\hat V^b{^\ddagger}\hat V^b}{\psi}=\braket*{\tilde{\psi}}{\psi}=\sum_n\abs{c_n}^2.
\end{equation}

In the following section we present the unitary implementation steps of Eq.\eqref{eq24} in a quantum computer.

 \subsection{The implementation algorithm}\label{sec:3.1}
We start with the  biorthgonal amplitude preparation $a_k\to\frac{1}{c}c_k$ in the $\mathcal{N}$ qubit state $\ket{\Psi}$, performed by a unitary operator $\hat{U}_{prep}$,
\begin{equation}\label{eq27}
\ket{\Bar{\psi}}=\hat{U}_{prep}\ket{\Psi}=\frac{1}{c}\sum_kc_k\ket{k},
\end{equation}
with
\begin{equation}\label{eq28}
c=\sqrt{\sum_k \abs{c_k}^2}.
\end{equation}
The implementation cost of the unitary operator $\hat{U}_T$ will be ignored, treated as an oracle operation.

Coupling the state $\ket{\Bar{\psi}}$ in Eq.\eqref{eq27} with an $\mathcal{N}$ qubit environment in the zero state $\ket{0}^{\otimes\mathcal{N}}$, we apply the $\hat{\mathcal{V}}\otimes \hat{1}$ operator where the unitary $\hat{\mathcal{V}}$ operator is defined in Eq.\eqref{eq26},
\begin{equation}\label{eq29}
(\hat{\mathcal{V}}\otimes \hat{1})(\ket{\Bar{\psi}}\otimes\ket{0}^{\otimes\mathcal{N}})=\frac{1}{c}\sum_{n,m}V_{nm}c_m\ket{n}\otimes\ket{0}^{\otimes\mathit{n}}.
\end{equation}

Next, apply at most $2^\mathcal{N}$ consecutive $n$-fold controlled operators $C^n\hat{U}_n$ in the ancillary state, $\hat{U}_n\ket{0}^{\otimes\mathcal{N}}=\ket{u_n}$, to generate the non-orthogonal basis $\{\ket{\bol u}\}$,
\begin{equation}\label{eq30}
C^n\hat{U}_n(\hat{\mathcal{V}}\otimes \hat{1})(\ket{\Bar{\psi}}\otimes\ket{0}^{\otimes\mathcal{N}})=\frac{1}{c}\sum_{n,m}\ket{n}\otimes V_{nm}c_m\ket{u_n}.
\end{equation}
Finally, a Quantum Fourier Transform (QFT) \cite{Nielsen_2010} in the first register will produce the desired state $\hat{V}\ket{\Psi}$ together with an orthogonal state $\ket{\perp}$,
\begin{equation}\label{eq31}
\begin{split}
&(QFT\otimes\hat{1})C^n\hat{U}_n(\hat{\mathcal{V}}\otimes \hat{1})(\ket{\Bar{\psi}}\otimes\ket{0}^{\otimes\mathcal{N}})=\\
&=\ket{0}^{\otimes\mathcal{N}}\otimes\frac{1}{c\sqrt{2^{\mathcal{N}}}}\hat{V}\ket{\Psi}+\ket{\perp},
\end{split}
\end{equation}
with $\bra*{\perp}(\ket{0}^{\otimes\mathcal{N}}\otimes\frac{1}{c\sqrt{2^{\mathcal{N}}}}\hat{V}\ket{\Psi})=0$.

Thus, a projective measurement $\hat{\Pi}_0=\ket{0}^{\otimes\mathcal{N}} \prescript{{\mathcal{N}}\otimes}{}{\bra{0}}\otimes\hat{1}$ on the first register,implements the action of the non-unitary operator $\hat{V}$ in the quantum state $\ket{\Psi}$ up a normalization factor,
\begin{equation}\label{eq32}
\ket{\Psi}_{out}=\frac{\hat{V}\ket{\Psi}}{\sqrt{\norm*{\hat{V}\ket{\Psi}}}},
\end{equation}
with success probability $p_{success}$,
\begin{equation}\label{eq33}
p_{success}=\frac{\mel{\Psi}{\hat{V}^\dagger\hat{V}}{\Psi}}{c^22^{\mathcal{N}}}.
\end{equation}

\begin{figure}[ht]
    \centering
    \includegraphics[width=\linewidth]{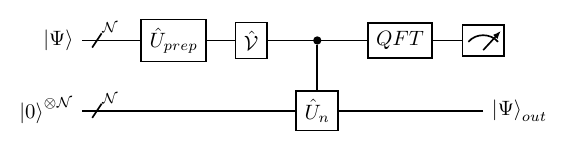}
    \caption{The quantum circuit implementation of quantum operations \eqref{eq27}-\eqref{eq33} that produce the output state $\ket{\Psi}_{out}\propto\hat{V}\ket{\Psi}$. The controlled operation represents a series of $2^{\mathcal{N}}$ multi-qubit controlled operations.}
    \label{fig:1}
\end{figure}

The steps involved in the implementation process described in Eqs.\eqref{eq27}-\eqref{eq32} are illustrated in Fig.\ref{fig:1}.

\subsection{A simple example}\label{sec:3.2}
Suppose we want to implement the single qubit non-unitary operator
\begin{equation}\label{eq34}
\hat{V}=\hat{\sigma}_z-(\hat{\sigma}_x+i\sigma_y)=\begin{bmatrix}
1&-2\\
0&-1
\end{bmatrix},
\end{equation}
in the $\{\ket{0}, \ket{1}\}$ computational basis with $\hat{\sigma}_i$ being the Pauli matrices and  $\hat{V}\ket{\Psi}=(a_0-2a_1)\ket{0}-a_1\ket{1}$.  

Selecting the following biorthogonal basis
\begin{alignat}{2}
&\ket{u_0}=\ket{0},\quad  && \ket{\zeta_0}=\ket{0}-\ket{1},\label{eq35}\\
&\ket{u_1}=\frac{\ket{0}+\ket{1}}{\sqrt{2}},\quad && \ket{\zeta_1}=\sqrt{2}\ket{1} \label{eq36},
\end{alignat}
the biorthogonal expression of non-unitary operator $\hat{V}$ becomes bi-unitary,
\begin{equation}\label{eq37}
\hat{V}^b=\ket{u_0}\bra{\zeta_0}-\ket{u_1}\bra{\zeta_1},\quad \hat{V}^{b^\ddagger}\hat{V}^b=\hat{1}^b,
\end{equation}
mimicking the action of the $\hat{\sigma}_z$ Pauli matrix in the biorthogonal basis and therefore $\hat{\mathcal{V}}=\hat{\sigma}_z$. The coefficients $c_0, c_1$ for the $\ket{\psi}$ state in the $\{\ket{u_0}, \ket{u_1}\}$ basis representation are $c_0=a_0-a_1$ and $c_1=\sqrt{2}a_1$.

Thus, following the procedure delineated in Sec.\ref{sec:3.1} we obtain,
\begin{equation}\label{eq38}
\begin{split}
\ket{step1}&=\hat{U}_{prep}\ket{\Psi}=\frac{1}{c}(c_0\ket{0}+c_1\ket{1}),\\
\ket{step2}&=(\hat{\sigma}_z\otimes\hat{1})(\ket{step1}\otimes\ket{0})\\
&=\frac{1}{c}(c_0\ket{0}-c_1\ket{1})\otimes\ket{0},\\
\ket{step3}&=C\hat{H}\ket{step2}\\
&=\frac{1}{c}(c_0\ket{0}\otimes\ket{u_0}-c_1\ket{1}\otimes\ket{u_1}),\\
\ket{step4}&=(\hat{H}\otimes\hat{1})\ket{step3}\\
&=\ket{0}\otimes\frac{1}{c\sqrt{2}}\hat{V}^b\ket{\psi}+\ket{\perp},\\
&\to\hat{V}^b\ket{\psi}=\hat{V}\ket{\Psi}
\end{split}
\end{equation}
The last step  in Eq.\eqref{eq38} incorporates a $0$-bit measurement in the first register with the normalization factor in the output state being omitted. In addition, the $\hat{H}$ gate is the Hadamard gate and the $C\hat{H}$ is the controlled  Hadamard gate,
\begin{equation}\label{eq39}
\hat{H}=\frac{1}{\sqrt{2}}\begin{bmatrix}
1&1\\
1&-1
\end{bmatrix},\quad C\hat{H}=\ket{0}\bra{0}\otimes\hat{1}+\ket{1}\bra{1}\otimes\hat{H}
\end{equation}
The success probability $p_{success}$ of process is,
\begin{equation}\label{eq40}
p_{success}=\frac{\abs{a_0-2a_1}^2+\abs{a_1}^2}{2(\abs{a_0-a_1}^2+2\abs{a_1}^2)}.
\end{equation}

\begin{figure}[ht]
    \centering
    \includegraphics[width=\linewidth]{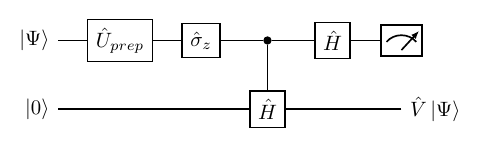}
    \caption{Quantum circuit implementation of non-unitary operator $\hat{V}$ in Eq.\eqref{eq34}, following the procedure of Eq.\eqref{eq38}. The measurement outcome in the first register is conditioned on the  $0$-bit value.}
    \label{fig:2}
\end{figure}

The quantum circuit implementation of the operations in Eq.\eqref{eq38} is depicted in Fig.\ref{fig:2}.

\subsection{Relaxing the bi-unitarity condition for non-unitary operators beyond contractions}\label{sec:3.3}
By allowing $\kappa_n>0$, the bi-unitarity condition in Eq.\eqref{eq21} along with the structure of the unity element $\hat{1}^b$ in Eq.\eqref{eq13} implies that every bi-unitary operator $\hat{V}^b$ serves as a biorthogonal unitarization for a class of biorthogonal operators $\hat{V}_c^b$ that satisfy,
\begin{align}
&\hat{V}_c^b=\sum_{n,m}V^c_{nm}\ket{u_n}\bra{\zeta_m},\label{eq42}\\
&\sum_k V^{c^*}_{nk}V^c_{km}=\delta_{nm},\label{eq43}\\ 
&\hat{V}^{b^\ddagger}_c\hat{V}^b_c=\sum_n \kappa_n\ket{u_n}\bra{\zeta_n}\label{eq44}.
\end{align}
Equation \eqref{eq43} dictates that the matrix representation $V^c_{nm}$ is unitary and subsequently the corresponding operator $\hat{\mathcal{V}}_c$ in Eq.\eqref{eqextra1} is unitary in the orthonormal basis. Therefore the implementation method proposed in Sec.\ref{sec:3.1} remains applicable, with an adjustment in the transformation of the biorthogonal amplitudes $c_n\to\kappa_nc_n$,
\begin{equation}\label{eq45}
\hat{U}_{prep}\ket{\Psi}=\sum_n \kappa_n c_n\ket{n},\quad c=\sqrt{\sum_n \kappa_n^2\abs{c_n}^2}.
\end{equation}

As a result, the requirement for implementing a non-unitary operator $\hat{V}$ to have a bi-unitary counterpart $\hat{V}^b$ is relaxed to the condition that  $\hat{V}^b$must have a unitary matrix representation in the biorthogonal basis. While this relaxation might initially appear trivial, it enables our method to implement non-unitary operators with diverse stability characteristics, accommodating both amplifying and dissipative effects. 

To illustrate our point, consider a parametrization of the biorthogonal basis in Eqs.\eqref{eq35} and \eqref{eq36},
\begin{alignat}{3}
&\ket{u_0}=\ket{0},\quad  && \ket{\zeta_0}=\tau(\ket{0}-\ket{1}),&&\quad \kappa_0=\tau,\\
&\ket{u_1}=\frac{\ket{0}+\ket{1}}{\sqrt{2}},\quad && \ket{\zeta_1}=\sqrt{2}\ket{1},&&\quad \kappa_1=1,
\end{alignat}
with  parameter $\tau>0$. In contrast to the case of Sec.\ref{sec:3.2} now the operator $\hat{V^b}$ is non bi-unitary for every $\tau>0$,
\begin{equation}
\hat{V}^b=\ket{u_0}\bra{\zeta_0}-\ket{u_1}\bra{\zeta_1},\quad \hat{V}^{b^\ddagger}\hat{V}^b\neq\hat{1}^b.
\end{equation}

However, $\hat{V}^b$ fulfils the conditions \eqref{eq42}-\eqref{eq44} for every $\tau>0$ and therefore the non-unitary operator $\hat{V}$,
\begin{equation}\label{eq49}
\hat{V}=\begin{bmatrix}
\tau & -(\tau+1)\\
0& -1
\end{bmatrix},
\end{equation}
is implementable within our framework.

For $\tau<1$ the eigenvalues $\lambda$ of matrix $\hat{V}$ are $\lambda=\{\tau<1, -1\}$. Therefore, the non-unitary matrix is a contraction. When $\tau=1$, the matrix reduces to the non-unitary case discussed in Sec.\ref{sec:3.2}, which possesses a bi-unitary representation. Most importantly,for $\tau>1$ the eigenvalues become $\lambda=\{\tau>1, -1\}$ corresponding to an amplification component. In all three cases,  the same algorithm procure of Sec.\ref{sec:3.1} can be applied with the general amplitude transformation in Eq.\eqref{eq45}.

\subsection{Implementation scaling and comparison with the LCU and Sz.-Nagy techniques}\label{sec:3.4}
The algorithm for implementing the non-unitary operator $\hat{V}$, along with its generalization as presented in Secs.\ref{sec:3.1} and \ref{sec:3.3} poses as a unitary $\mathcal{N}$-qubit dilation method. Naturally, we have to compare its advantages and limitations against other methods for implementing non-unitary operations, namely the Sz.Nagy dilation \cite{Moshe_2021} and the LCU method \cite{Childs_2012}.

Our dilation consists of a quantum Fourier transform that can be implemented using $\textit{O}(\mathcal{N}^2)$ elementary gates as well as $\textit{O}(2^{\mathcal{N}})$ $n$-fold controlled $\hat{U}_n$ gates that admits an overall implementation within $\textit{O}(\mathcal{N}^2 2^{\mathcal{N}})$ single qubit and Controlled-NOT (CNOT) gates. Additionally, assuming that the unitary operator $\hat{\mathcal{V}}$ can be  efficiently decomposed in $\textit{O}[poly(\mathcal{N})]$ simple gates, the total implementation cost scales as $\textit{O}(\mathcal{N}^2 2^{\mathcal{N}})$. Note that the implementation cost of the unitary oracle operation $\hat{U}_{prep}$ has not been considered in the previous analysis. 

An interesting attribute of our algorithm is that the success probability $p_{success}$ decreases with the dimension of the the non-unitary matrix $\hat{V}$, as $p_{success}\sim 1/2^{\mathcal{N}}$. However, this probability remains unchanged regardless of the number of summands of unitary operations into which $\hat{V}$ is decomposed.  This is a striking difference with the LCU method where the success probability scales as $p_{sucess}^{LCU}\sim 1/2^{{N}}$, where $N$ is the number of the unitary summands. The LCU method also requires $\log_2{N}$ ancillary qubits. As a result for non-unitary contraction operators, $\norm*{\hat{V}}<1$, both the LCU and the Sz. Nagy techniques outperform our method. The LCU method offers a significantly higher success probability, while the Sz.-Nagy dilation is deterministic, both being  one-qubit $(N=2)$ dilation methods.

The situation changes drastically when attempting to implement a non-contraction non-unitary operator $\hat{V}$ such as the one in Eq.\eqref{eq49} for $\tau\geq 1$. In such cases the Sz. Nagy dilation is not applicable since the defect operator $\hat{D}_V$ \cite{Moshe_2021},
\begin{equation}\label{eq50}
\hat{D}_V=\sqrt{\hat{1}-\hat{V}^\dagger\hat{V}}
\end{equation}
becomes ill defined. The number $N$ of unitary summands in the linear decomposition of a $2^{\mathcal{N}}$-dimensional non-contraction operator $\hat{V}$ is $N=\mathit{O}(4^{\mathcal{N}})$, rendering the LCU method ineffective due to a very low success probability and the required $2\mathcal{N}$ ancillary qubits. In contrast, our method is independent of the unitary decomposition and thus clearly prevails above both the Sz.-Nagy and LCU methods in cases of non-contractions, offering higher success probability and requiring fewer ancillary qubits.

To showcase the advantages of the biorthogonal dilation method for implementing non-contractions, we utilize the LCU method \cite{Childs_2012} to implement the marginal case, $\tau=1$, non-unitary matrix of Eq.\eqref{eq34} in order to draw explicit comparisons with the biorthogonal implementation of Sec.\ref{sec:3.2}. Take notice that because the operator $\hat{V}^\dagger\hat{V}$ has an eigenvalue $\lambda=3+2\sqrt{2}$, the defect operator of the Sz.-Nagy dilation in Eq.\eqref{eq50} fails to be implemented. To apply the LCU method, we first re-write the unitary decomposition in Eq.\eqref{eq34} using positive coefficients,
\begin{equation}\label{eq51}
\hat{V}=\hat{\sigma}_z+ (-\hat{\sigma}_x)+\hat{\sigma}_x\hat{\sigma}_z=\begin{bmatrix}
1&-2\\
0&-1
\end{bmatrix}.
\end{equation}

Immediately, since there are thee unitary summands, $\hat{U}_1=\hat{\sigma}_z$, $\hat{U}_1=-\hat{\sigma}_x$ and $\hat{U}_3=\hat{\sigma}_x\hat{\sigma}_z$, two ancillary qubits are required. The respective preparation operator $\hat{U}_{prep}^{LCU}$ acts on the ancillary register as 
\begin{equation}\label{eq52}
\hat{U}_{prep}^{LCU}\ket{0}=\frac{1}{\sqrt{3}}(\ket{0}+\ket{1}+\ket{2}).
\end{equation}
Defining, the following unitary two-qubit control gates,
\begin{equation}\label{eq53}
C^i\hat{U}_i=\ket{i}\bra{i}\otimes\hat{U}_i,\quad i=1,2,3,
\end{equation}
the implementation sequence reads,
\begin{equation}\label{eq54}
\hat{U}_{prep}^{{LCU}^\dagger}\prod_{i}(C^i\hat{U}_i)\hat{U}_{prep}^{LCU}(\ket{0}\otimes\ket{\Psi})=\frac{1}{\sqrt{3}}\ket{0}\otimes\hat{V}\ket{\Psi}+\ket{\perp}.
\end{equation}
Therefore, according to Eq.\eqref{eq54}, the success implementation probability of $\hat{V}$ is,
\begin{equation}\label{eq55}
p_{sucess}^{LCU}=\frac{\abs{a_0-2a_1}^2+\abs{a_1}^2}{3}.
\end{equation}

\begin{figure}[ht]
    \centering
    \includegraphics[width=\linewidth]{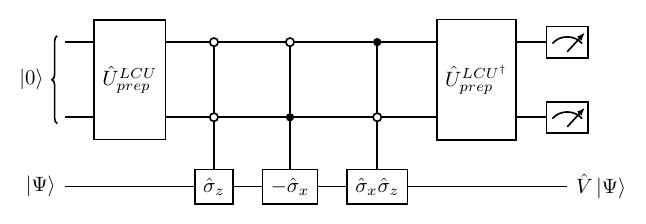}
    \caption{The LCU quantum circuit of Eq.\eqref{eq54}, implementing the non-unitary operator $\hat{V}$ in Eq.\eqref{eq34}.}
    \label{fig:3}
\end{figure}

The LCU success probability in Eq.\eqref{eq55} and the biorthogonal success probability in Eq.\eqref{eq40} for the implementation of the same non-unitary operator $\hat{V}$ (Eq.\eqref{eq34}) are comparable since $c^2\sim 1$. However, the LCU accomplishes the implementation of the non-unitary operator as a two-qubit dilation, while the biorthogonal dilation accomplishes the same task with only one extra qubit. Evidently, by comparing the implementation quantum circuits illustrated in Fig.\ref{fig:3} for the LCU and  Fig.\ref{fig:2} for the biorthogonal dilation, the LCU circuit has significantly deeper depth. This includes two invokes to the oracle operation $\hat{U}_{prep}^{LCU}$ as well as three two-qubit controlled gates.

\subsection{Discussion on potential applications}\label{sec:3.5}
Taking into consideration the implementation scaling and the advantages of the biorthogonal dilation as presented in Sec.\ref{sec:3.4}, we promote a synergistic application of both LCU and dilation methods for efficient quantum implementation of large dimensional non-unitary operators possessing a small-dimensional rank subspace of eigenvalues $\lambda$ with $\abs{\lambda}>1$. In this scenario, the LCU protocol efficiently handles the large-dimensional contraction space with $\abs{\lambda}<1$, whereas the present biorthogonal dilation implementation process addresses the few amplification components, $\abs{\lambda}>1$. 

The aforementioned hybrid approach is well suited for implementation of non-unitary operations in positive only and possibly trace-preserving (PTP) quantum systems. These systems typically arise occur when the initial correlations between the open system $\rho_S$ and the environment $\rho_E$ cannot be ignored, leading to a non-separable composite density matrix $\rho_{SE}\neq\rho_S\otimes\rho_E$. Then, the reduced dynamics of the open system
\begin{equation}\label{eq56}
\rho'_S=\mathcal{E}(\rho_S)=Tr_E(\hat{U}_{SE}\rho_{SE}\hat{U}^\dagger_{SE}),
\end{equation}
extends beyond the Markovian completely positive and trace-preserving (CPTP) Kraus representation \cite{Pechukas_1994,Stelmachovic_2001,Shaji_2005,Breuer_2007,Breuer_2016,Rembielinski_2021,Schlimgen2_2022}. Thus, the reduced dynamics can be non-linear and PTP. In Eq.\eqref{eq56}, $\hat{U}_{SE}$ is the unitary operator acting on the composite system. For example, in the linear PTP quantum channels,
\begin{equation}\label{eq57}
\rho'_S=\mathcal{E}(\rho_S)=\sum_j\gamma_j\hat{K}_j\rho_S\hat{K}^\dagger_j, \quad \gamma_j=\pm 1,
\end{equation}
for some of the participating Kraus operators $\hat{K}_j$ applies that $\norm*{\hat{K}_j}>1$. Hence, a quantum computing implementation of Eq.\eqref{eq57} is not possible under the Sz.-Nagy dilation of the Kraus operators as in \cite{Hu_2020}. Instead, it could be facilitated by  the proposed LCU--biorthogonal approach.

Another potential application is in the pseudo-Hermitian systems \cite{Mostafazadeh_2002}, which exhibit either a purely real spectrum or a complex conjugates eigenvalues, marking the transition from stable dynamics to instabilities and dissipation in classical physical systems \cite{Qin_2019,Zhang_2020}. As such, the dissipative part in the associated non-unitary dynamics can be treated with the LCU while the amplifying part can be managed with the biorthogonal method. In quantum systems the break of a pseudo-Hermitian symmetry is related with the violation of adiabicity \cite{Sim_2023}. As such, instead of solving the time dependent pseudo-Hermicity relation for the time-dependent metric operator $\bol{\hat{f}}(t)$ \cite{Znozil_2008,Fring_2016} one can use the biorthogonal framework \cite{Hou_2024} and consequently the biorthogonal dilation method for quantum implementation of the non-adiabatic dynamics.

\section{Conclusions}\label{sec:4}
The present paper aims to establish a $\mathcal{N}$-qubit  dilation protocol for the quantum implementation of a $2^{\mathcal{N}}$-dimensional non-unitary operation for which the LCU method has proven to encounter challenges. The implementation cost of the method scales as $\textit{O}(\mathcal{N}^2 2^{\mathcal{N}})$ using a single unitary oracle.

The algorithm only requires that the non-unitary operator to possess a biorthogonal matrix representation that is unitary. Albeit this requirement may initially seem restrictive, such a construction is fairly general \cite{Dieudonne_1953}. The characteristics of the algorithm are complementary to those of the LCU method, with the implementation success probability  to decrease with the dimensionality of the non-unitary operator but to remain unaffected by the number of unitary summands, regardless of the stability properties of the non-unitary operator.

Therefore, when the non-unitary operator of interest is not a contraction, the biorthogonal dilation method proves more efficient than the LCU method and can be leveraged as a sub-routine in the quantum implementation of a general non-unitary operators possessing a small dimensional component with eigenvalues $\abs{\lambda}>1$. Such dynamics are present in realistic physical systems in both  quantum and classical realms, notably in PTP open quantum systems and pseudo-Hermitian systems.\\

\section*{Acknowledgments}
This work has been carried out within the framework of the EUROfusion Consortium, funded by the European Union via the Euratom Research and Training Programme (Grant Agreement No 101052200 — EUROfusion). Views and opinions expressed are however those of the authors only and do not necessarily reflect those of the European Union or the European Commission. Neither the European Union nor the European Commission can be held responsible for them.
A.K.R is supported by the US Department of Energy under Grant Nos. DE-SC0021647 and DE-FG02-91ER-54109.
G.V is supported by the US Department of Energy under Grant No. DE-SC0021651.

\bibliography{ref}

\providecommand{\noopsort}[1]{}\providecommand{\singleletter}[1]{#1}%
\begin{thebibliography}{36}%
\makeatletter
\providecommand \@ifxundefined [1]{%
 \@ifx{#1\undefined}
}%
\providecommand \@ifnum [1]{%
 \ifnum #1\expandafter \@firstoftwo
 \else \expandafter \@secondoftwo
 \fi
}%
\providecommand \@ifx [1]{%
 \ifx #1\expandafter \@firstoftwo
 \else \expandafter \@secondoftwo
 \fi
}%
\providecommand \natexlab [1]{#1}%
\providecommand \enquote  [1]{``#1''}%
\providecommand \bibnamefont  [1]{#1}%
\providecommand \bibfnamefont [1]{#1}%
\providecommand \citenamefont [1]{#1}%
\providecommand \href@noop [0]{\@secondoftwo}%
\providecommand \href [0]{\begingroup \@sanitize@url \@href}%
\providecommand \@href[1]{\@@startlink{#1}\@@href}%
\providecommand \@@href[1]{\endgroup#1\@@endlink}%
\providecommand \@sanitize@url [0]{\catcode `\\12\catcode `\$12\catcode `\&12\catcode `\#12\catcode `\^12\catcode `\_12\catcode `\%12\relax}%
\providecommand \@@startlink[1]{}%
\providecommand \@@endlink[0]{}%
\providecommand \url  [0]{\begingroup\@sanitize@url \@url }%
\providecommand \@url [1]{\endgroup\@href {#1}{\urlprefix }}%
\providecommand \urlprefix  [0]{URL }%
\providecommand \Eprint [0]{\href }%
\providecommand \doibase [0]{https://doi.org/}%
\providecommand \selectlanguage [0]{\@gobble}%
\providecommand \bibinfo  [0]{\@secondoftwo}%
\providecommand \bibfield  [0]{\@secondoftwo}%
\providecommand \translation [1]{[#1]}%
\providecommand \BibitemOpen [0]{}%
\providecommand \bibitemStop [0]{}%
\providecommand \bibitemNoStop [0]{.\EOS\space}%
\providecommand \EOS [0]{\spacefactor3000\relax}%
\providecommand \BibitemShut  [1]{\csname bibitem#1\endcsname}%
\let\auto@bib@innerbib\@empty
%</preamble>
\bibitem [{\citenamefont {Breuer}\ and\ \citenamefont {Petruccione}(2007)}]{Breuer_2007}%
  \BibitemOpen
  \bibfield  {author} {\bibinfo {author} {\bibfnamefont {H.-P.}\ \bibnamefont {Breuer}}\ and\ \bibinfo {author} {\bibfnamefont {F.}~\bibnamefont {Petruccione}},\ }\href {https://doi.org/10.1093/acprof:oso/9780199213900.001.0001} {\emph {\bibinfo {title} {{The Theory of Open Quantum Systems}}}}\ (\bibinfo  {publisher} {Oxford University Press},\ \bibinfo {year} {2007})\BibitemShut {NoStop}%
\bibitem [{\citenamefont {Delgado-Granados}\ \emph {et~al.}(2024)\citenamefont {Delgado-Granados}, \citenamefont {Krogmeier}, \citenamefont {Sager-Smith}, \citenamefont {Avdic}, \citenamefont {Hu}, \citenamefont {Sajjan}, \citenamefont {Abbasi}, \citenamefont {Smart}, \citenamefont {Narang}, \citenamefont {Kais}, \citenamefont {Schlimgen}, \citenamefont {Head-Marsden},\ and\ \citenamefont {Mazziotti}}]{Delgado_2024}%
  \BibitemOpen
  \bibfield  {author} {\bibinfo {author} {\bibfnamefont {L.~H.}\ \bibnamefont {Delgado-Granados}}, \bibinfo {author} {\bibfnamefont {T.~J.}\ \bibnamefont {Krogmeier}}, \bibinfo {author} {\bibfnamefont {L.~M.}\ \bibnamefont {Sager-Smith}}, \bibinfo {author} {\bibfnamefont {I.}~\bibnamefont {Avdic}}, \bibinfo {author} {\bibfnamefont {Z.}~\bibnamefont {Hu}}, \bibinfo {author} {\bibfnamefont {M.}~\bibnamefont {Sajjan}}, \bibinfo {author} {\bibfnamefont {M.}~\bibnamefont {Abbasi}}, \bibinfo {author} {\bibfnamefont {S.~E.}\ \bibnamefont {Smart}}, \bibinfo {author} {\bibfnamefont {P.}~\bibnamefont {Narang}}, \bibinfo {author} {\bibfnamefont {S.}~\bibnamefont {Kais}}, \bibinfo {author} {\bibfnamefont {A.~W.}\ \bibnamefont {Schlimgen}}, \bibinfo {author} {\bibfnamefont {K.}~\bibnamefont {Head-Marsden}},\ and\ \bibinfo {author} {\bibfnamefont {D.~A.}\ \bibnamefont {Mazziotti}},\ }\href {https://doi.org/10.48550/arXiv.2406.05219} {\bibinfo {title} {Quantum algorithms and applications for open quantum systems}} (\bibinfo
  {year} {2024}),\ \Eprint {https://arxiv.org/abs/2406.05219} {arXiv:2406.05219 [quant-ph]} \BibitemShut {NoStop}%
\bibitem [{\citenamefont {Mostafazadeh}(2010)}]{Mostafazadeh_2010}%
  \BibitemOpen
  \bibfield  {author} {\bibinfo {author} {\bibfnamefont {A.}~\bibnamefont {Mostafazadeh}},\ }\bibfield  {title} {\bibinfo {title} {Pseudo-hermitian representation of quantum mechanics},\ }\href {https://doi.org/10.1142/S0219887810004816} {\bibfield  {journal} {\bibinfo  {journal} {International Journal of Geometric Methods in Modern Physics}\ }\textbf {\bibinfo {volume} {07}},\ \bibinfo {pages} {1191} (\bibinfo {year} {2010})}\BibitemShut {NoStop}%
\bibitem [{\citenamefont {Ju}\ \emph {et~al.}(2019)\citenamefont {Ju}, \citenamefont {Miranowicz}, \citenamefont {Chen},\ and\ \citenamefont {Nori}}]{Ju_2019}%
  \BibitemOpen
  \bibfield  {author} {\bibinfo {author} {\bibfnamefont {C.-Y.}\ \bibnamefont {Ju}}, \bibinfo {author} {\bibfnamefont {A.}~\bibnamefont {Miranowicz}}, \bibinfo {author} {\bibfnamefont {G.-Y.}\ \bibnamefont {Chen}},\ and\ \bibinfo {author} {\bibfnamefont {F.}~\bibnamefont {Nori}},\ }\bibfield  {title} {\bibinfo {title} {Non-hermitian hamiltonians and no-go theorems in quantum information},\ }\href {https://doi.org/10.1103/PhysRevA.100.062118} {\bibfield  {journal} {\bibinfo  {journal} {Phys. Rev. A}\ }\textbf {\bibinfo {volume} {100}},\ \bibinfo {pages} {062118} (\bibinfo {year} {2019})}\BibitemShut {NoStop}%
\bibitem [{\citenamefont {Childs}\ and\ \citenamefont {Wiebe}(2012)}]{Childs_2012}%
  \BibitemOpen
  \bibfield  {author} {\bibinfo {author} {\bibfnamefont {A.~M.}\ \bibnamefont {Childs}}\ and\ \bibinfo {author} {\bibfnamefont {N.}~\bibnamefont {Wiebe}},\ }\bibfield  {title} {\bibinfo {title} {Hamiltonian simulation using linear combinations of unitary operations},\ }\href@noop {} {\bibfield  {journal} {\bibinfo  {journal} {Quantum Info. Comput.}\ }\textbf {\bibinfo {volume} {12}},\ \bibinfo {pages} {901–924} (\bibinfo {year} {2012})}\BibitemShut {NoStop}%
\bibitem [{\citenamefont {Berry}\ \emph {et~al.}(2015)\citenamefont {Berry}, \citenamefont {Childs},\ and\ \citenamefont {Kothari}}]{Berry_2015}%
  \BibitemOpen
  \bibfield  {author} {\bibinfo {author} {\bibfnamefont {D.~W.}\ \bibnamefont {Berry}}, \bibinfo {author} {\bibfnamefont {A.~M.}\ \bibnamefont {Childs}},\ and\ \bibinfo {author} {\bibfnamefont {R.}~\bibnamefont {Kothari}},\ }\bibfield  {title} {\bibinfo {title} {Hamiltonian simulation with nearly optimal dependence on all parameters},\ }in\ \href {https://doi.org/10.1109/FOCS.2015.54} {\emph {\bibinfo {booktitle} {Proceedings of the 2015 IEEE 56th Annual Symposium on Foundations of Computer Science (FOCS)}}},\ \bibinfo {series and number} {FOCS '15}\ (\bibinfo {year} {2015})\ p.\ \bibinfo {pages} {792–809}\BibitemShut {NoStop}%
\bibitem [{\citenamefont {Krovi}(2023)}]{Krovi_2023}%
  \BibitemOpen
  \bibfield  {author} {\bibinfo {author} {\bibfnamefont {H.}~\bibnamefont {Krovi}},\ }\bibfield  {title} {\bibinfo {title} {Improved quantum algorithms for linear and nonlinear differential equations},\ }\href {https://doi.org/10.22331/q-2023-02-02-913} {\bibfield  {journal} {\bibinfo  {journal} {{Quantum}}\ }\textbf {\bibinfo {volume} {7}},\ \bibinfo {pages} {913} (\bibinfo {year} {2023})}\BibitemShut {NoStop}%
\bibitem [{\citenamefont {Gui-Lu}(2006)}]{Long_2006}%
  \BibitemOpen
  \bibfield  {author} {\bibinfo {author} {\bibfnamefont {L.}~\bibnamefont {Gui-Lu}},\ }\bibfield  {title} {\bibinfo {title} {General quantum interference principle and duality computer},\ }\href {https://doi.org/10.1088/0253-6102/45/5/013} {\bibfield  {journal} {\bibinfo  {journal} {Commun. Theor. Phys.}\ }\textbf {\bibinfo {volume} {45}},\ \bibinfo {pages} {825} (\bibinfo {year} {2006})}\BibitemShut {NoStop}%
\bibitem [{\citenamefont {Zheng}(2021)}]{Zheng_2021}%
  \BibitemOpen
  \bibfield  {author} {\bibinfo {author} {\bibfnamefont {C.}~\bibnamefont {Zheng}},\ }\bibfield  {title} {\bibinfo {title} {{Universal quantum simulation of single-qubit nonunitary operators using duality quantum algorithm}},\ }\href {https://doi.org/10.1038/s41598-021-83521-5} {\bibfield  {journal} {\bibinfo  {journal} {Sci Rep}\ }\textbf {\bibinfo {volume} {11}},\ \bibinfo {pages} {3960} (\bibinfo {year} {2021})}\BibitemShut {NoStop}%
\bibitem [{\citenamefont {Schlimgen}\ \emph {et~al.}(2021)\citenamefont {Schlimgen}, \citenamefont {Head-Marsden}, \citenamefont {Sager}, \citenamefont {Narang},\ and\ \citenamefont {Mazziotti}}]{Schlimgen_2021}%
  \BibitemOpen
  \bibfield  {author} {\bibinfo {author} {\bibfnamefont {A.~W.}\ \bibnamefont {Schlimgen}}, \bibinfo {author} {\bibfnamefont {K.}~\bibnamefont {Head-Marsden}}, \bibinfo {author} {\bibfnamefont {L.~M.}\ \bibnamefont {Sager}}, \bibinfo {author} {\bibfnamefont {P.}~\bibnamefont {Narang}},\ and\ \bibinfo {author} {\bibfnamefont {D.~A.}\ \bibnamefont {Mazziotti}},\ }\bibfield  {title} {\bibinfo {title} {Quantum simulation of open quantum systems using a unitary decomposition of operators},\ }\href {https://doi.org/10.1103/PhysRevLett.127.270503} {\bibfield  {journal} {\bibinfo  {journal} {Phys. Rev. Lett.}\ }\textbf {\bibinfo {volume} {127}},\ \bibinfo {pages} {270503} (\bibinfo {year} {2021})}\BibitemShut {NoStop}%
\bibitem [{\citenamefont {Suri}\ \emph {et~al.}(2023)\citenamefont {Suri}, \citenamefont {Barreto}, \citenamefont {Hadfield}, \citenamefont {Wiebe}, \citenamefont {Wudarski},\ and\ \citenamefont {Marshall}}]{Suri_2023}%
  \BibitemOpen
  \bibfield  {author} {\bibinfo {author} {\bibfnamefont {N.}~\bibnamefont {Suri}}, \bibinfo {author} {\bibfnamefont {J.}~\bibnamefont {Barreto}}, \bibinfo {author} {\bibfnamefont {S.}~\bibnamefont {Hadfield}}, \bibinfo {author} {\bibfnamefont {N.}~\bibnamefont {Wiebe}}, \bibinfo {author} {\bibfnamefont {F.}~\bibnamefont {Wudarski}},\ and\ \bibinfo {author} {\bibfnamefont {J.}~\bibnamefont {Marshall}},\ }\bibfield  {title} {\bibinfo {title} {Two-{U}nitary {D}ecomposition {A}lgorithm and {O}pen {Q}uantum {S}ystem {S}imulation},\ }\href {https://doi.org/10.22331/q-2023-05-15-1002} {\bibfield  {journal} {\bibinfo  {journal} {{Quantum}}\ }\textbf {\bibinfo {volume} {7}},\ \bibinfo {pages} {1002} (\bibinfo {year} {2023})}\BibitemShut {NoStop}%
\bibitem [{\citenamefont {Schlimgen}\ \emph {et~al.}(2022{\natexlab{a}})\citenamefont {Schlimgen}, \citenamefont {Head-Marsden}, \citenamefont {Sager-Smith}, \citenamefont {Narang},\ and\ \citenamefont {Mazziotti}}]{Schlimgen_2022}%
  \BibitemOpen
  \bibfield  {author} {\bibinfo {author} {\bibfnamefont {A.~W.}\ \bibnamefont {Schlimgen}}, \bibinfo {author} {\bibfnamefont {K.}~\bibnamefont {Head-Marsden}}, \bibinfo {author} {\bibfnamefont {L.~M.}\ \bibnamefont {Sager-Smith}}, \bibinfo {author} {\bibfnamefont {P.}~\bibnamefont {Narang}},\ and\ \bibinfo {author} {\bibfnamefont {D.~A.}\ \bibnamefont {Mazziotti}},\ }\bibfield  {title} {\bibinfo {title} {Quantum state preparation and nonunitary evolution with diagonal operators},\ }\href {https://doi.org/10.1103/PhysRevA.106.022414} {\bibfield  {journal} {\bibinfo  {journal} {Phys. Rev. A}\ }\textbf {\bibinfo {volume} {106}},\ \bibinfo {pages} {022414} (\bibinfo {year} {2022}{\natexlab{a}})}\BibitemShut {NoStop}%
\bibitem [{\citenamefont {Schlimgen}\ \emph {et~al.}(2022{\natexlab{b}})\citenamefont {Schlimgen}, \citenamefont {Head-Marsden}, \citenamefont {Sager-Smith}, \citenamefont {Narang},\ and\ \citenamefont {Mazziotti}}]{Schlimgen2_2022}%
  \BibitemOpen
  \bibfield  {author} {\bibinfo {author} {\bibfnamefont {A.~W.}\ \bibnamefont {Schlimgen}}, \bibinfo {author} {\bibfnamefont {K.}~\bibnamefont {Head-Marsden}}, \bibinfo {author} {\bibfnamefont {L.~M.}\ \bibnamefont {Sager-Smith}}, \bibinfo {author} {\bibfnamefont {P.}~\bibnamefont {Narang}},\ and\ \bibinfo {author} {\bibfnamefont {D.~A.}\ \bibnamefont {Mazziotti}},\ }\href {https://doi.org/10.48550/arXiv.2207.07112} {\bibinfo {title} {Quantum simulation of open quantum systems using density-matrix purification}} (\bibinfo {year} {2022}{\natexlab{b}}),\ \Eprint {https://arxiv.org/abs/2207.07112} {arXiv:2207.07112 [quant-ph]} \BibitemShut {NoStop}%
\bibitem [{\citenamefont {Jin}\ \emph {et~al.}(2023)\citenamefont {Jin}, \citenamefont {Liu},\ and\ \citenamefont {Yu}}]{Jin_2023}%
  \BibitemOpen
  \bibfield  {author} {\bibinfo {author} {\bibfnamefont {S.}~\bibnamefont {Jin}}, \bibinfo {author} {\bibfnamefont {N.}~\bibnamefont {Liu}},\ and\ \bibinfo {author} {\bibfnamefont {Y.}~\bibnamefont {Yu}},\ }\bibfield  {title} {\bibinfo {title} {Quantum simulation of partial differential equations: Applications and detailed analysis},\ }\href {https://doi.org/10.1103/PhysRevA.108.032603} {\bibfield  {journal} {\bibinfo  {journal} {Phys. Rev. A}\ }\textbf {\bibinfo {volume} {108}},\ \bibinfo {pages} {032603} (\bibinfo {year} {2023})}\BibitemShut {NoStop}%
\bibitem [{\citenamefont {Koukoutsis}\ \emph {et~al.}(2024)\citenamefont {Koukoutsis}, \citenamefont {Hizanidis}, \citenamefont {Ram},\ and\ \citenamefont {Vahala}}]{Koukoutsis_2024}%
  \BibitemOpen
  \bibfield  {author} {\bibinfo {author} {\bibfnamefont {E.}~\bibnamefont {Koukoutsis}}, \bibinfo {author} {\bibfnamefont {K.}~\bibnamefont {Hizanidis}}, \bibinfo {author} {\bibfnamefont {A.~K.}\ \bibnamefont {Ram}},\ and\ \bibinfo {author} {\bibfnamefont {G.}~\bibnamefont {Vahala}},\ }\bibfield  {title} {\bibinfo {title} {Quantum simulation of dissipation for maxwell equations in dispersive media},\ }\href {https://doi.org/https://doi.org/10.1016/j.future.2024.05.028} {\bibfield  {journal} {\bibinfo  {journal} {Future Gener. Comput. Syst.}\ }\textbf {\bibinfo {volume} {159}},\ \bibinfo {pages} {221} (\bibinfo {year} {2024})}\BibitemShut {NoStop}%
\bibitem [{\citenamefont {Shalit}(2021)}]{Moshe_2021}%
  \BibitemOpen
  \bibfield  {author} {\bibinfo {author} {\bibfnamefont {O.~M.}\ \bibnamefont {Shalit}},\ }\bibfield  {title} {\bibinfo {title} {Dilation theory: A guided tour},\ }in\ \href {https://doi.org/10.1007/978-3-030-51945-2_28} {\emph {\bibinfo {booktitle} {Operator Theory, Functional Analysis and Applications}}},\ \bibinfo {editor} {edited by\ \bibinfo {editor} {\bibfnamefont {M.~A.}\ \bibnamefont {Bastos}}, \bibinfo {editor} {\bibfnamefont {L.}~\bibnamefont {Castro}},\ and\ \bibinfo {editor} {\bibfnamefont {A.~Y.}\ \bibnamefont {Karlovich}}}\ (\bibinfo  {publisher} {Springer International Publishing},\ \bibinfo {year} {2021})\ pp.\ \bibinfo {pages} {551--623}\BibitemShut {NoStop}%
\bibitem [{\citenamefont {Brody}(2013)}]{Brody_2013}%
  \BibitemOpen
  \bibfield  {author} {\bibinfo {author} {\bibfnamefont {D.~C.}\ \bibnamefont {Brody}},\ }\bibfield  {title} {\bibinfo {title} {Biorthogonal quantum mechanics},\ }\href {https://doi.org/10.1088/1751-8113/47/3/035305} {\bibfield  {journal} {\bibinfo  {journal} {J. Phys. A}\ }\textbf {\bibinfo {volume} {47}},\ \bibinfo {pages} {035305} (\bibinfo {year} {2013})}\BibitemShut {NoStop}%
\bibitem [{\citenamefont {Mostafazadeh}(2002)}]{Mostafazadeh_2002}%
  \BibitemOpen
  \bibfield  {author} {\bibinfo {author} {\bibfnamefont {A.}~\bibnamefont {Mostafazadeh}},\ }\bibfield  {title} {\bibinfo {title} {{Pseudo-Hermiticity versus PT symmetry: The necessary condition for the reality of the spectrum of a non-Hermitian Hamiltonian}},\ }\href {https://doi.org/10.1063/1.1418246} {\bibfield  {journal} {\bibinfo  {journal} {J. Math. Phys.}\ }\textbf {\bibinfo {volume} {43}},\ \bibinfo {pages} {205} (\bibinfo {year} {2002})}\BibitemShut {NoStop}%
\bibitem [{\citenamefont {Znojil}(2008)}]{Znozil_2008}%
  \BibitemOpen
  \bibfield  {author} {\bibinfo {author} {\bibfnamefont {M.}~\bibnamefont {Znojil}},\ }\bibfield  {title} {\bibinfo {title} {Time-dependent version of crypto-hermitian quantum theory},\ }\href {https://doi.org/10.1103/PhysRevD.78.085003} {\bibfield  {journal} {\bibinfo  {journal} {Phys. Rev. D}\ }\textbf {\bibinfo {volume} {78}},\ \bibinfo {pages} {085003} (\bibinfo {year} {2008})}\BibitemShut {NoStop}%
\bibitem [{\citenamefont {Mostafazadeh}(2003)}]{Mostafazadeh_2003}%
  \BibitemOpen
  \bibfield  {author} {\bibinfo {author} {\bibfnamefont {A.}~\bibnamefont {Mostafazadeh}},\ }\bibfield  {title} {\bibinfo {title} {{Is Pseudo-Hermitian Quantum Mechanics an Indefinite-Metric Quantum Theory}},\ }\href {https://doi.org/10.1023/B:CJOP.0000010537.23790.8c} {\bibfield  {journal} {\bibinfo  {journal} {Czech. J. Phys.}\ }\textbf {\bibinfo {volume} {53}},\ \bibinfo {pages} {1079–1084} (\bibinfo {year} {2003})}\BibitemShut {NoStop}%
\bibitem [{\citenamefont {Koukoutsis}\ \emph {et~al.}(2023)\citenamefont {Koukoutsis}, \citenamefont {Hizanidis}, \citenamefont {Ram},\ and\ \citenamefont {Vahala}}]{Koukoutsis_2023}%
  \BibitemOpen
  \bibfield  {author} {\bibinfo {author} {\bibfnamefont {E.}~\bibnamefont {Koukoutsis}}, \bibinfo {author} {\bibfnamefont {K.}~\bibnamefont {Hizanidis}}, \bibinfo {author} {\bibfnamefont {A.~K.}\ \bibnamefont {Ram}},\ and\ \bibinfo {author} {\bibfnamefont {G.}~\bibnamefont {Vahala}},\ }\bibfield  {title} {\bibinfo {title} {Dyson maps and unitary evolution for maxwell equations in tensor dielectric media},\ }\href {https://doi.org/10.1103/PhysRevA.107.042215} {\bibfield  {journal} {\bibinfo  {journal} {Phys. Rev. A}\ }\textbf {\bibinfo {volume} {107}},\ \bibinfo {pages} {042215} (\bibinfo {year} {2023})}\BibitemShut {NoStop}%
\bibitem [{\citenamefont {Zhang}\ and\ \citenamefont {Wu}(2018)}]{Zhang_2018}%
  \BibitemOpen
  \bibfield  {author} {\bibinfo {author} {\bibfnamefont {Q.}~\bibnamefont {Zhang}}\ and\ \bibinfo {author} {\bibfnamefont {B.}~\bibnamefont {Wu}},\ }\bibfield  {title} {\bibinfo {title} {Lorentz quantum mechanics},\ }\href {https://doi.org/10.1088/1367-2630/aa8496} {\bibfield  {journal} {\bibinfo  {journal} {New J. Phys.}\ }\textbf {\bibinfo {volume} {20}},\ \bibinfo {pages} {013024} (\bibinfo {year} {2018})}\BibitemShut {NoStop}%
\bibitem [{\citenamefont {Edvardsson}\ \emph {et~al.}(2023)\citenamefont {Edvardsson}, \citenamefont {König},\ and\ \citenamefont {Stålhammar}}]{Edvardsson_2023}%
  \BibitemOpen
  \bibfield  {author} {\bibinfo {author} {\bibfnamefont {E.}~\bibnamefont {Edvardsson}}, \bibinfo {author} {\bibfnamefont {J.~L.~K.}\ \bibnamefont {König}},\ and\ \bibinfo {author} {\bibfnamefont {M.}~\bibnamefont {Stålhammar}},\ }\href {https://doi.org/10.48550/arXiv.2212.06004} {\bibinfo {title} {Biorthogonal renormalization}} (\bibinfo {year} {2023}),\ \Eprint {https://arxiv.org/abs/2212.06004} {arXiv:2212.06004 [quant-ph]} \BibitemShut {NoStop}%
\bibitem [{\citenamefont {Nielsen}\ and\ \citenamefont {Chuang}(2010)}]{Nielsen_2010}%
  \BibitemOpen
  \bibfield  {author} {\bibinfo {author} {\bibfnamefont {M.~A.}\ \bibnamefont {Nielsen}}\ and\ \bibinfo {author} {\bibfnamefont {I.~L.}\ \bibnamefont {Chuang}},\ }\href {https://doi.org/10.1017/CBO9780511976667} {\emph {\bibinfo {title} {Quantum Computation and Quantum Information: 10th Anniversary Edition}}}\ (\bibinfo  {publisher} {Cambridge University Press},\ \bibinfo {year} {2010})\BibitemShut {NoStop}%
\bibitem [{\citenamefont {Pechukas}(1994)}]{Pechukas_1994}%
  \BibitemOpen
  \bibfield  {author} {\bibinfo {author} {\bibfnamefont {P.}~\bibnamefont {Pechukas}},\ }\bibfield  {title} {\bibinfo {title} {Reduced dynamics need not be completely positive},\ }\href {https://doi.org/10.1103/PhysRevLett.73.1060} {\bibfield  {journal} {\bibinfo  {journal} {Phys. Rev. Lett.}\ }\textbf {\bibinfo {volume} {73}},\ \bibinfo {pages} {1060} (\bibinfo {year} {1994})}\BibitemShut {NoStop}%
\bibitem [{\citenamefont {\ifmmode \check{S}\else \v{S}\fi{}telmachovi\ifmmode~\check{c}\else \v{c}\fi{}}\ and\ \citenamefont {Bu\ifmmode~\check{z}\else \v{z}\fi{}ek}(2001)}]{Stelmachovic_2001}%
  \BibitemOpen
  \bibfield  {author} {\bibinfo {author} {\bibfnamefont {P.}~\bibnamefont {\ifmmode \check{S}\else \v{S}\fi{}telmachovi\ifmmode~\check{c}\else \v{c}\fi{}}}\ and\ \bibinfo {author} {\bibfnamefont {V.}~\bibnamefont {Bu\ifmmode~\check{z}\else \v{z}\fi{}ek}},\ }\bibfield  {title} {\bibinfo {title} {Dynamics of open quantum systems initially entangled with environment: Beyond the kraus representation},\ }\href {https://doi.org/10.1103/PhysRevA.64.062106} {\bibfield  {journal} {\bibinfo  {journal} {Phys. Rev. A}\ }\textbf {\bibinfo {volume} {64}},\ \bibinfo {pages} {062106} (\bibinfo {year} {2001})}\BibitemShut {NoStop}%
\bibitem [{\citenamefont {Shaji}\ and\ \citenamefont {Sudarshan}(2005)}]{Shaji_2005}%
  \BibitemOpen
  \bibfield  {author} {\bibinfo {author} {\bibfnamefont {A.}~\bibnamefont {Shaji}}\ and\ \bibinfo {author} {\bibfnamefont {E.}~\bibnamefont {Sudarshan}},\ }\bibfield  {title} {\bibinfo {title} {Who's afraid of not completely positive maps?},\ }\href {https://doi.org/10.1016/j.physleta.2005.04.029} {\bibfield  {journal} {\bibinfo  {journal} {Phys. Lett. A}\ }\textbf {\bibinfo {volume} {341}},\ \bibinfo {pages} {48} (\bibinfo {year} {2005})}\BibitemShut {NoStop}%
\bibitem [{\citenamefont {Breuer}\ \emph {et~al.}(2016)\citenamefont {Breuer}, \citenamefont {Laine}, \citenamefont {Piilo},\ and\ \citenamefont {Vacchini}}]{Breuer_2016}%
  \BibitemOpen
  \bibfield  {author} {\bibinfo {author} {\bibfnamefont {H.-P.}\ \bibnamefont {Breuer}}, \bibinfo {author} {\bibfnamefont {E.-M.}\ \bibnamefont {Laine}}, \bibinfo {author} {\bibfnamefont {J.}~\bibnamefont {Piilo}},\ and\ \bibinfo {author} {\bibfnamefont {B.}~\bibnamefont {Vacchini}},\ }\bibfield  {title} {\bibinfo {title} {Colloquium: Non-markovian dynamics in open quantum systems},\ }\href {https://doi.org/10.1103/RevModPhys.88.021002} {\bibfield  {journal} {\bibinfo  {journal} {Rev. Mod. Phys.}\ }\textbf {\bibinfo {volume} {88}},\ \bibinfo {pages} {021002} (\bibinfo {year} {2016})}\BibitemShut {NoStop}%
\bibitem [{\citenamefont {Rembieli{\'{n}}ski}\ and\ \citenamefont {Caban}(2021)}]{Rembielinski_2021}%
  \BibitemOpen
  \bibfield  {author} {\bibinfo {author} {\bibfnamefont {J.}~\bibnamefont {Rembieli{\'{n}}ski}}\ and\ \bibinfo {author} {\bibfnamefont {P.}~\bibnamefont {Caban}},\ }\bibfield  {title} {\bibinfo {title} {Nonlinear extension of the quantum dynamical semigroup},\ }\href {https://doi.org/10.22331/q-2021-03-23-420} {\bibfield  {journal} {\bibinfo  {journal} {{Quantum}}\ }\textbf {\bibinfo {volume} {5}},\ \bibinfo {pages} {420} (\bibinfo {year} {2021})}\BibitemShut {NoStop}%
\bibitem [{\citenamefont {Hu}\ and\ \citenamefont {Kais}(2020)}]{Hu_2020}%
  \BibitemOpen
  \bibfield  {author} {\bibinfo {author} {\bibfnamefont {R.}~\bibnamefont {Hu}, \bibfnamefont {Z.and~Xia}}\ and\ \bibinfo {author} {\bibfnamefont {S.}~\bibnamefont {Kais}},\ }\bibfield  {title} {\bibinfo {title} {A quantum algorithm for evolving open quantum dynamics on quantum computing devices},\ }\href {https://doi.org/10.1038/s41598-020-60321-x} {\bibfield  {journal} {\bibinfo  {journal} {Sci. Rep.}\ }\textbf {\bibinfo {volume} {10}},\ \bibinfo {pages} {3301} (\bibinfo {year} {2020})}\BibitemShut {NoStop}%
\bibitem [{\citenamefont {Qin}\ \emph {et~al.}(2019)\citenamefont {Qin}, \citenamefont {Zhang}, \citenamefont {Glasser},\ and\ \citenamefont {Xiao}}]{Qin_2019}%
  \BibitemOpen
  \bibfield  {author} {\bibinfo {author} {\bibfnamefont {H.}~\bibnamefont {Qin}}, \bibinfo {author} {\bibfnamefont {R.}~\bibnamefont {Zhang}}, \bibinfo {author} {\bibfnamefont {A.~S.}\ \bibnamefont {Glasser}},\ and\ \bibinfo {author} {\bibfnamefont {J.}~\bibnamefont {Xiao}},\ }\bibfield  {title} {\bibinfo {title} {{Kelvin-Helmholtz instability is the result of parity-time symmetry breaking}},\ }\href {https://doi.org/10.1063/1.5088498} {\bibfield  {journal} {\bibinfo  {journal} {Phys. Plasmas}\ }\textbf {\bibinfo {volume} {26}},\ \bibinfo {pages} {032102} (\bibinfo {year} {2019})}\BibitemShut {NoStop}%
\bibitem [{\citenamefont {Zhang}\ \emph {et~al.}(2020)\citenamefont {Zhang}, \citenamefont {Qin},\ and\ \citenamefont {Xiao}}]{Zhang_2020}%
  \BibitemOpen
  \bibfield  {author} {\bibinfo {author} {\bibfnamefont {R.}~\bibnamefont {Zhang}}, \bibinfo {author} {\bibfnamefont {H.}~\bibnamefont {Qin}},\ and\ \bibinfo {author} {\bibfnamefont {J.}~\bibnamefont {Xiao}},\ }\bibfield  {title} {\bibinfo {title} {{PT-symmetry entails pseudo-Hermiticity regardless of diagonalizability}},\ }\href {https://doi.org/10.1063/1.5117211} {\bibfield  {journal} {\bibinfo  {journal} {J. Math. Phys.}\ }\textbf {\bibinfo {volume} {61}},\ \bibinfo {pages} {012101} (\bibinfo {year} {2020})}\BibitemShut {NoStop}%
\bibitem [{\citenamefont {Sim}\ \emph {et~al.}(2023)\citenamefont {Sim}, \citenamefont {Defenu}, \citenamefont {Molignini},\ and\ \citenamefont {Chitra}}]{Sim_2023}%
  \BibitemOpen
  \bibfield  {author} {\bibinfo {author} {\bibfnamefont {K.}~\bibnamefont {Sim}}, \bibinfo {author} {\bibfnamefont {N.}~\bibnamefont {Defenu}}, \bibinfo {author} {\bibfnamefont {P.}~\bibnamefont {Molignini}},\ and\ \bibinfo {author} {\bibfnamefont {R.}~\bibnamefont {Chitra}},\ }\bibfield  {title} {\bibinfo {title} {Quantum metric unveils defect freezing in non-hermitian systems},\ }\href {https://doi.org/10.1103/PhysRevLett.131.156501} {\bibfield  {journal} {\bibinfo  {journal} {Phys. Rev. Lett.}\ }\textbf {\bibinfo {volume} {131}},\ \bibinfo {pages} {156501} (\bibinfo {year} {2023})}\BibitemShut {NoStop}%
\bibitem [{\citenamefont {Fring}\ and\ \citenamefont {Moussa}(2016)}]{Fring_2016}%
  \BibitemOpen
  \bibfield  {author} {\bibinfo {author} {\bibfnamefont {A.}~\bibnamefont {Fring}}\ and\ \bibinfo {author} {\bibfnamefont {M.~H.~Y.}\ \bibnamefont {Moussa}},\ }\bibfield  {title} {\bibinfo {title} {Unitary quantum evolution for time-dependent quasi-hermitian systems with nonobservable hamiltonians},\ }\href {https://doi.org/10.1103/PhysRevA.93.042114} {\bibfield  {journal} {\bibinfo  {journal} {Phys. Rev. A}\ }\textbf {\bibinfo {volume} {93}},\ \bibinfo {pages} {042114} (\bibinfo {year} {2016})}\BibitemShut {NoStop}%
\bibitem [{\citenamefont {Hou}\ and\ \citenamefont {Li}(2024)}]{Hou_2024}%
  \BibitemOpen
  \bibfield  {author} {\bibinfo {author} {\bibfnamefont {T.-X.}\ \bibnamefont {Hou}}\ and\ \bibinfo {author} {\bibfnamefont {W.}~\bibnamefont {Li}},\ }\bibfield  {title} {\bibinfo {title} {Nonadiabatic geometric quantum computation in non-hermitian systems},\ }\href {https://doi.org/10.1103/PhysRevA.109.022616} {\bibfield  {journal} {\bibinfo  {journal} {Phys. Rev. A}\ }\textbf {\bibinfo {volume} {109}},\ \bibinfo {pages} {022616} (\bibinfo {year} {2024})}\BibitemShut {NoStop}%
\bibitem [{\citenamefont {Dieudonn{\'e}}(1953)}]{Dieudonne_1953}%
  \BibitemOpen
  \bibfield  {author} {\bibinfo {author} {\bibfnamefont {J.}~\bibnamefont {Dieudonn{\'e}}},\ }\bibfield  {title} {\bibinfo {title} {On biorthogonal systems},\ }\href {https://doi.org/10.1307/mmj/1028989861} {\bibfield  {journal} {\bibinfo  {journal} {Mich. Math. J.}\ }\textbf {\bibinfo {volume} {2}},\ \bibinfo {pages} {7} (\bibinfo {year} {1953})}\BibitemShut {NoStop}%
\end{thebibliography}%

\end{document}